\documentclass[structabstract]{aa}  

\usepackage{graphicx}
\usepackage{txfonts}
\usepackage{natbib}
\bibpunct{(}{)}{;}{a}{}{,}
\usepackage{multirow}
\usepackage{nicefrac}
\usepackage[rightcaption]{sidecap}
\usepackage[usenames,dvipsnames]{color}
\usepackage{hyperref}
\hypersetup{
    final=true,
    pageanchor=true,
    colorlinks=true,
    breaklinks=true,
    linkcolor=blue,
    citecolor=blue,
    urlcolor=blue,
    pdfpagemode=UseNone}

\begin{document}


\title{A giant quiescent solar filament observed with high-resolution spectroscopy}

\author{C.\ Kuckein
    \and M.\ Verma 
    \and C.\ Denker
}   

\institute{Leibniz-Institut f\"ur Astrophysik Potsdam (AIP), 
    An der Sternwarte 16, 
    14482 Potsdam, Germany\\
    \email{ckuckein@aip.de}}

\date{Version: \today} 

\abstract
{}
{An extremely large filament was studied in various layers of the solar atmosphere. The inferred
physical parameters and the morphological aspects are compared with smaller quiescent filaments. } 
{A giant, quiet-Sun filament was observed with the high-resolution Echelle
spectrograph at the Vacuum Tower Telescope at Observatorio del Teide, Tenerife,
Spain on 2011 November 15. A mosaic of spectra (10 maps of $100\arcsec \times
182\arcsec$) was recorded simultaneously in the chromospheric absorption lines H$\alpha$
and \ion{Na}{i} D$_2$. Physical parameters of the filament plasma were derived using
\textit{Cloud Model} (CM) inversions and line core fits. The spectra were complemented with full-disk
filtergrams (\ion{He}{i} $\lambda$10830~\AA, H$\alpha$, and \ion{Ca}{ii}\,K) of
the Chromspheric Telescope (ChroTel) and full-disk magnetograms of the Helioseismic and
Magnetic Imager (HMI).}
{The filament had extremely large linear dimensions ($\sim$817 arcsec), which 
corresponds to about 658~Mm along a great circle on the solar surface. A total amount of 175119 H$\alpha$ contrast 
profiles were inverted using the CM approach. The inferred mean line-of-sight (LOS) velocity, Doppler width, and source 
function were similar to previous works of smaller quiescent filaments. However, the derived optical thickness was 
larger. LOS velocity trends inferred from the H$\alpha$ line core fits were in accord, but smaller, than the ones 
obtained with CM inversions. Signatures of counter-streaming flows were detected in the filament. 
The largest conglomerates of 
brightenings in the line core of \ion{Na}{i} D$_2$ coincided well with small-scale magnetic fields as seen by HMI. 
Mixed magnetic polarities were detected close to the ends of barbs. The computation of photospheric horizontal flows 
based on HMI magnetograms revealed flow kernels with a size of 5--8~Mm and velocities of 0.30--0.45~km~s$^{-1}$ at 
the ends of the filament.}
{The physical properties of extremely large filaments are similar to their smaller counterparts, except for the optical 
thickness which in our sample was found to be larger. We found that a part of the filament, which erupted the day 
before, is in the process of reestablishing its initial configuration.}


\authorrunning{Kuckein et al.}

\titlerunning{A giant quiescent solar filament observed with high-resolution spectroscopy}

\maketitle


\section{Introduction}

Filaments are still a puzzling topic in solar physics. Although many studies have been
carried out in the past decades, new high-resolution observations open the door to a
better understanding of this common solar phenomenon. Filaments are located in the corona with their lower parts 
in the chromosphere. They appear as elongated dark features on the disk
whereas above the limb, they shine bright against the dark background of space and are
called prominences.
However, both terms are often interchangeably used in literature. We can distinguish between three types of 
filaments: quiescent, intermediate, and active region (AR) filaments \citep[e.g.,][]{engvold15}. The main difference among 
them relies on the magnetic flux density found around the filament. Active region filaments are found in areas of high 
magnetic flux density. In comparison, quiescent filaments are found outside of ARs, have much larger
dimensions, and lie higher in the atmosphere \citep[e.g.,][]{aulanier03,lites05} than AR filaments. 
The term intermediate filaments is used when no clear classification into quiescent or AR is possible.

In this study, we concentrate on quiescent filaments, which have been
extensively studied in previous works by other authors. For a complete review of 
filaments the reader is referred to \citet{tandberg95}, \citet{mackay10}, 
\citet{parenti14}, and \citet{vial15}. In almost all filaments an elongated main axis, 
better known as the spine, is identified. In addition, small structures, which are called barbs,
protrude from the spine towards both sides. Barbs appear to be rooted in 
the photosphere -- among parasitic polarities \citep[e.g.,][]{martin98,chae05,lopezariste06}.

The typical length of filaments is between 60 and 600~Mm \citep[][]{tandberg95}. Nevertheless,
larger filaments exist (> 600~Mm) but are scarcely discussed in literature. 
In one of the few instances, \citet[][]{anderson05} showed that extraordinary long filament channels, which evolve 
into filaments, are formed by merging of smaller ones. The authors observed a channel that reached a maximum length of 
160 heliographic degrees along the northern hemisphere. 
Whether the resulting filament, after merging, presents the same 
physical properties as the former smaller individual filaments is still unknown. \citet[][]{yazev88} reported on a 
giant 
filament with an extension of up to 800~Mm. This particular
aspect of extreme linear dimensions is the main driver for the present work. 

Filaments are supported against gravity by magnetic field lines.
Previous works favor a twisted magnetic field configuration 
\citep[e.g.,][]{vanballe89,devore00,guo10,kuckein12a,yelles12}, although non-twisted field scenarios also exist
\citep[e.g.,][]{kippen57}. Following this argument, a giant helical structure can support an extremely
large filament in the solar atmosphere. This structure is anchored tens of megameters below in the photosphere. 
The question which naturally arises is how stable is such a filament. Quiescent filaments are much 
more stable than their smaller AR counterparts and can survive for several weeks. Nevertheless, 
emerging and submerging flows, as well as shearing motions close to the polarity inversion line,
are responsible for the destabilization of the filament which often leads to flares or coronal 
mass ejections. As a consequence, a partial or complete eruption of the filament can happen.

\begin{figure}[t]
\includegraphics[width=\columnwidth]{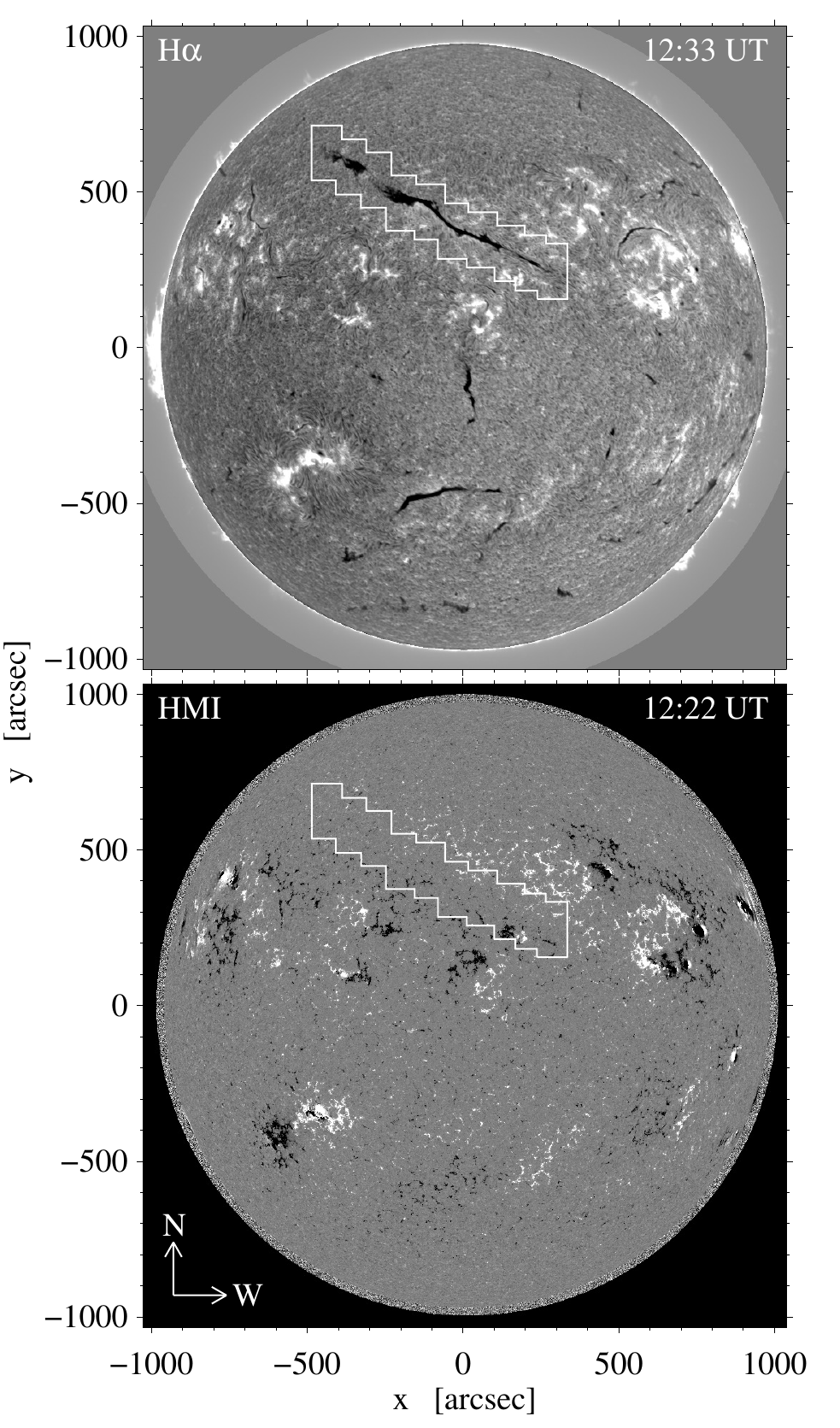}
\caption{Limb-darkening corrected H$\alpha$ full-disk image obtained with
    ChroTel on 2011 November 15 (\textit{top}). Full-disk SDO/HMI magnetogram scaled between
    $\pm$100\,G (\textit{bottom}). The rectangular contours outline the regions
    scanned with the Echelle spectrograph of the VTT.}
\label{Fig:ChroHMI}
\end{figure}

\begin{figure}[t]
\includegraphics[width=\columnwidth]{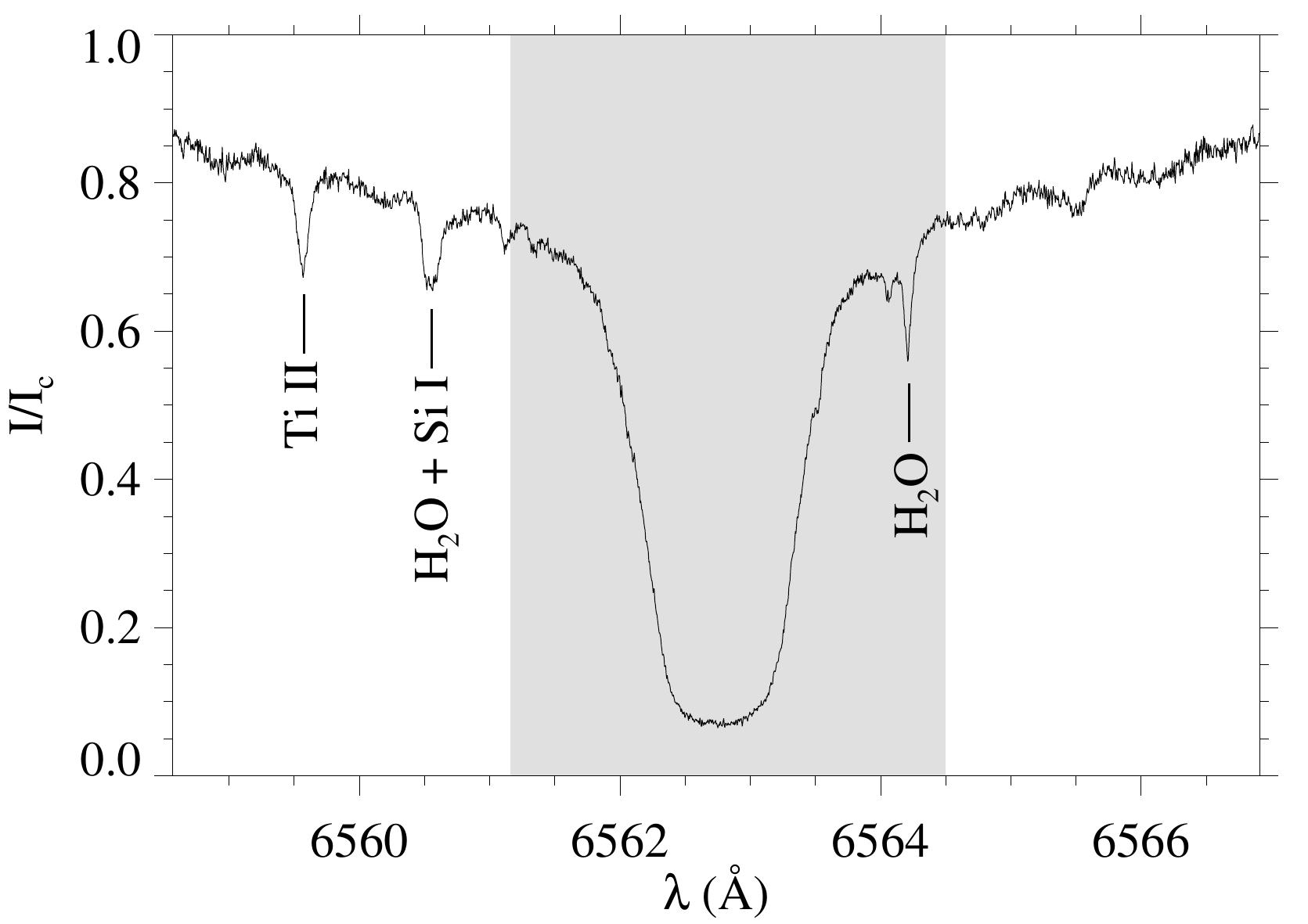}
\caption{H$\alpha$ profile acquired with the Echelle spectrograph at the VTT. The profile belongs to the spine of the
filament and is located at label `b' in Fig. \ref{Fig:vttmaps}. The shaded area represents the spectral range used to 
compute the contrast profiles.}
\label{Fig:Haprofile}
\end{figure}

The by far most used spectral line to study the morphology of filaments is the
chromospheric H$\alpha$ $\lambda$6562.8~\AA\ line. It reveals the smallest fine
structure detected in filaments \citep[e.g.,][]{lin05,heinzel06} and
provides information about the chirality and mass flows \citep[e.g.,][]{martin98}. 
Doppler shifts of the H$\alpha$ line have been the quintessence to retrieve 
line-of-sight (LOS) velocities in filaments. \citet{martres1981} found that blueshifts ($< 4$~km\,s$^{-1}$) were
long-lasting and localized in dark structures. Furthermore, high velocities (7~km\,s$^{-1}$) were detected in
transient events. Based on the observation that the highest H$\alpha$ velocities
were correlated to high-brightness horizontal gradients, they concluded that the
filament and surrounding bright regions are part of one dynamical and geometrical
structure. Dynamical H$\alpha$ fine structures of a quiescent filament were the topic under study by
\citet{mein1994}. The filament is formed by an accumulation of tiny thin threads. 
The authors modeled these H$\alpha$ threads, and the best fit
between observed and simulated quantities was obtained for an optical
thickness of $\tau_{0} = 0.2$, a source function of $S_{0}=0.06$, and a mean upward velocity of
$v_{0} = 1.7$~km\,s$^{-1}$. Observations of the H$\alpha$ line in the red and blue wing revealed counter-streaming 
flows, i.e., simultaneous flows in opposite directions, in the spine and barbs of filaments \citep[][]{zirker98}. 
Velocities of individual knots of mass were in the range of 5\,--\,20 ~km\,s$^{-1}$. Nevertheless, much higher 
velocities were detected when tracking individual knots observed in \ion{He}{ii} 304~\AA\ along a quiescent prominence. 
The typical inferred values were $\sim$30~km\,s$^{-1}$ with peaks up to 75~km\,s$^{-1}$ \citep[][]{wang99}.

A well-known method to infer 
physical parameters from the H$\alpha$ line is the Cloud Model (CM) inversion technique developed by
\citet[][]{beckers64}. This relatively simple model assumes a cloud of material suspended 
above the photosphere (the basic assumptions are discussed in Sect. \ref{Sect:cloudmodelinv}). 
Various properties such as the optical thickness, LOS velocity, Doppler width, and source function
can be obtained from the inversion of the contrast profiles derived from the H$\alpha$ line. 
For instance, \citet{maltby1976}  
applied CM inversions to filament features. 
The author found up and downflows on the order of few km\,s$^{-1}$ in the filament.  
Using the first-order differential cloud model (DCM), \citet{schmieder1991} detected 
high velocities of up to $\pm 15$~km\,s$^{-1}$ at the footpoints of the filament. 
Beckers's CM technique was used by \citet{chae06} to infer the physical parameters of a quiescent
filament observed at the Big Bear Solar Observatory. Two-dimensional maps of the parameters were computed.
If the contrast was high enough, the H$\alpha$
contrast profiles were correctly fitted. The authors concluded that the LOS velocities
retrieved by CM inversions are much more reasonable than the ones obtained by the line core fits. 

It has not been studied if there is a relation between the optical thickness and the length of the 
filament. \citet[][]{chae06} and \citet[][]{schmieder03} used CM inversions to retrieve the 
optical thickness in relatively small filaments compared to the one studied in this work. The authors reported on 
optical thickness values of $0.83\pm0.47$ and $0.66$, respectively.

Improved CM versions have been used to examine other 
chromospheric features including fibrils \citep[e.g.,][]{bray1974}, 
arch filaments \citep[e.g.,][]{alissandrakis90},
mottles \citep[e.g.,][]{bray1973}, and surges \citep[e.g.,][]{gu1994}.
\citet{tzio07} reviewed the different CM inversion techniques and their
application to various chromospheric features.

Another chromospheric line, the \ion{Na}{i} D$_2$ line at $\lambda$5890~\AA, has been used 
for plasma diagnostics in prominences.
For instance, \citet{landman83} computed line ratios of this line and other lines 
to derive physical parameters such as the prominence electron
density. The sodium line is optically thin in prominences and is usually studied
together with the neighboring \ion{Na}{i} D$_1$ line at $\lambda$5896~\AA\ \citep{landman81}.
Observations revealed that most prominences with significant
emission in \ion{Na}{i} D$_2$ and Mg b had pronounced centrally reversed H$\alpha$
profiles \citep[][]{stellmacher2005}.

\begin{SCfigure*}[0.75][t]
\includegraphics[width=0.75\textwidth]{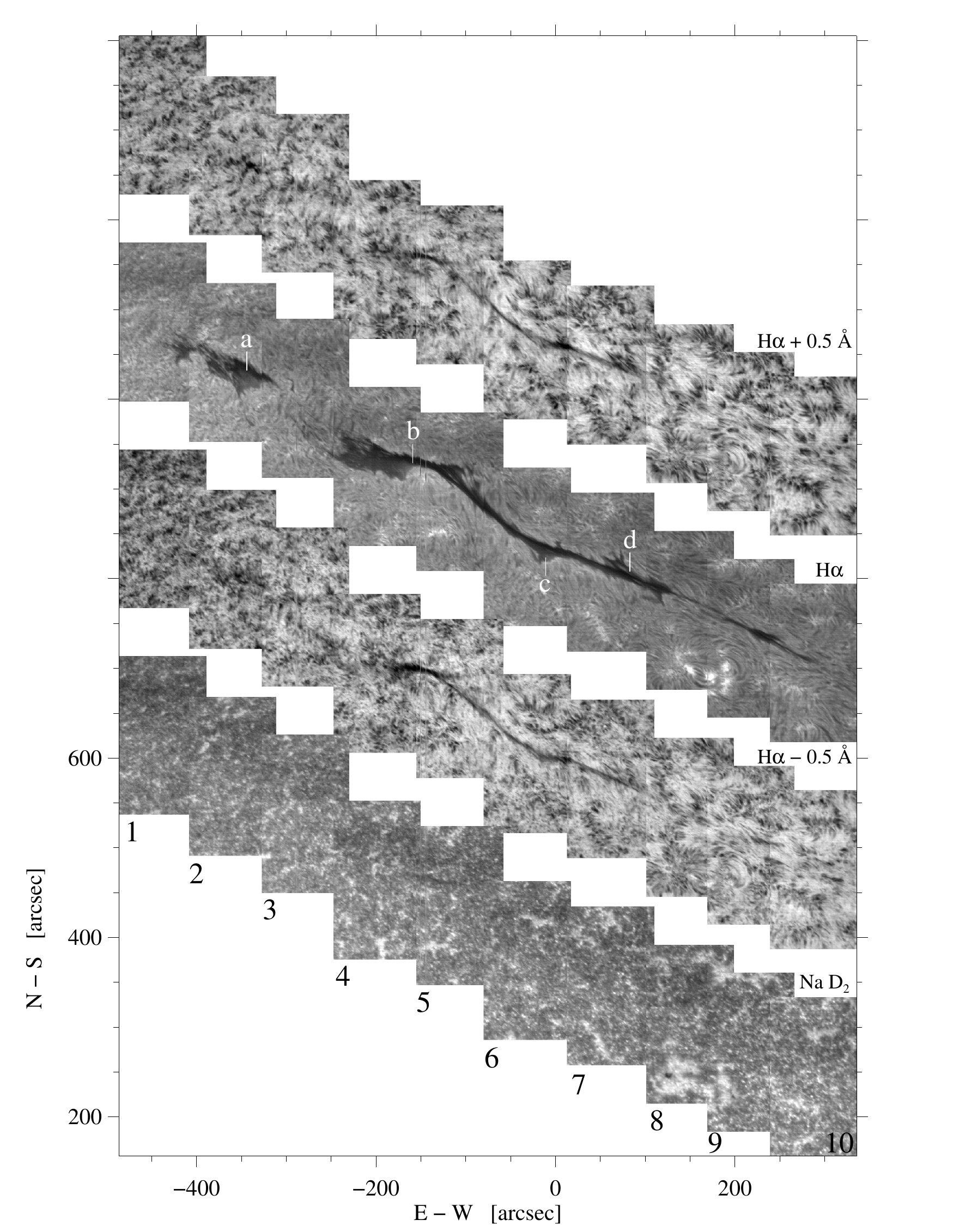}
\hspace*{-10mm}\vspace{-3mm}
\caption{The large filament is represented as a mosaic of ten individual 
    slit-reconstructed images centered at \mbox{H$\alpha+0.5$~\AA}, H$\alpha$
    line core, \mbox{H$\alpha-0.5$~\AA}, and \ion{Na}{i} D$_2$ line core (\textit{top} to
    \textit{bottom}). We enumerate the maps from left to right (1--10).   
    The labels `a'\,--\,`d' indicate the locations of the contrast
    profiles shown in Fig.~\ref{Fig:cprofiles}.}
\label{Fig:vttmaps}
\end{SCfigure*}

In summary, there have been numerous studies looking at the plasma properties
of quiescent filaments either by observations or modeling. The motivation of this work is to 
examine these properties in an uncommonly large quiescent filament which covered half of the solar disk based on 
simultaneous observations in the H$\alpha$ and
\ion{Na}{i} D$_2$ spectral lines (see Fig.~\ref{Fig:ChroHMI}). 
Preliminary results were presented by 
\citet[][]{kuckein14}. In Sect.~\ref{Sect:Results} we will focus on the results of CM inversions and
additionally compute the Doppler shifts to compare both methods in the case of the H$\alpha$ line. 
Furthermore, using magnetograms we will identify magnetic features below the barbs of the filament and additionally 
compute photospheric horizontal motions in search for converging flows.


\section{Observations}

\begin{SCfigure*}[0.68][t]
\includegraphics[width=0.68\textwidth]{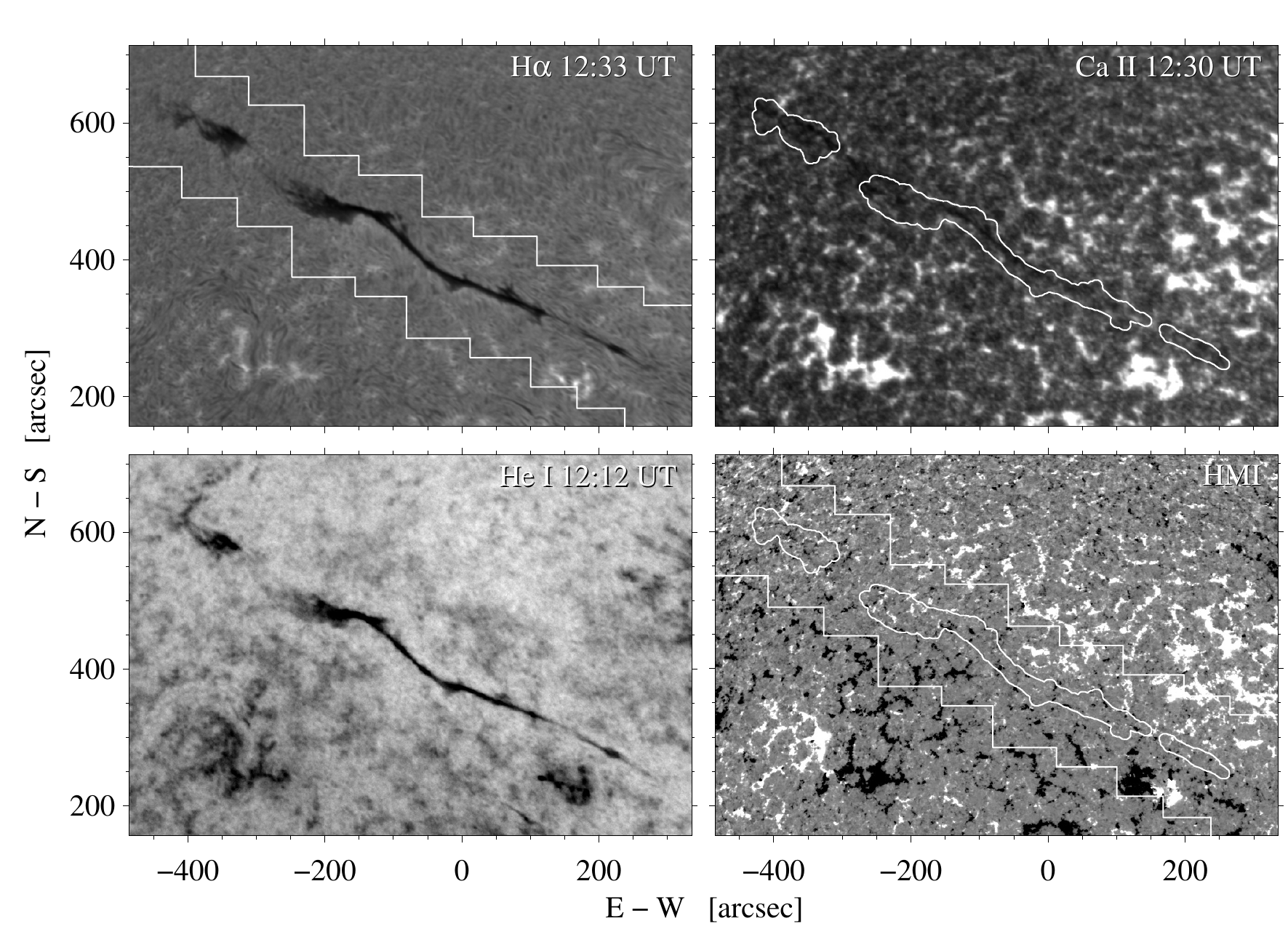}
\hspace*{3mm}\vspace{0mm}
\caption{Region-of-interest cropped from ChroTel and HMI full-disk images
    around 12:30~UT showing H$\alpha$, \ion{Ca}{ii}\,K, and \ion{He}{i}
    line core intensity images as well as a time-averaged (11:30--13:20~UT) HMI
    magnetogram (\textit{top-left} to \textit{bottom-right}). The magnetogram is clipped
    between $\pm$30~G. The step-function-like contours in the H$\alpha$ and HMI
    panels outline the mosaic of the Echelle spectrograph observations (see
    Fig.~\ref{Fig:vttmaps}). Intensity thresholding of the H$\alpha$ filament
    and subsequent morphological dilation was used to create the generous
    contours of the filament in the \ion{Ca}{ii}\,K and HMI panels.}
\label{Fig:chrotelmaps}
\end{SCfigure*}

The giant filament was already present on the backside of the Sun before
rotating to the front side. The leading extremity of the filament became visible
between 2011 November 8--9 at the eastern solar limb. The filament was
continuously present during its disk passage until November 22, when it
lifted off as part of a coronal mass ejection (CME) at the western solar limb.
The dynamical aspects of this filament were discussed in \citet[][]{diercke2014} and
a peer-reviewed publication is forthcoming. 

The high-resolution ground-based observations of the giant quiescent filament
were acquired on 2011 November 15 with the Echelle spectrograph of the German
Vacuum Tower Telescope \citep[VTT,][]{vonderLuehe1998} at the Observatorio del
Teide, Tenerife, Spain. The very good seeing conditions facilitated scanning the
whole filament from East to West across the northern hemisphere of the Sun. The
observing strategy was to divide the filament into ten pieces with a scanned
area of $100\arcsec \times 182\arcsec$ each. Consecutive scans with a duration of $\sim$10~min 
overlapped assuring
the continuity of the filament in the reconstructed mosaic. The scanning device
is an integral part of the Kiepenheuer-Institute Adaptive Optics System
\citep[KAOS,][]{berkefeld2010} at the VTT. The lock-point of KAOS was
in the center of the scanned area, and the whole filament was scanned piecewise
with a spatial step of 0.32\arcsec\ between 11:38 and 13:20~UT. The extreme
points of the filament were located at heliographic coordinates ($35^\circ$~East,
$37^\circ$~North) and ($18^\circ$~West, $11^\circ$~North). The filament's length
was roughly 817\arcsec.

Two PCO.4000 CCD cameras were mounted at the Echelle spectrograph to record
simultaneously spectra in two different spectral regions: (1) the chromospheric H$\alpha$ line
at $\lambda$6562.8~\AA\ within a spectral range of 8~\AA\ (see Fig. \ref{Fig:Haprofile}) and (2) the \ion{Na}{i} D$_2$
line at $\lambda$5889.9~\AA\ within a spectral range of 7~\AA. The spectral
sampling was 4.2 and 3.6~m\AA~pixel$^{-1}$, respectively. The pixel size along
the slit was 0.16\arcsec, which was half the spatial step when scanning the solar
surface. Exposure times were 600~ms. 
The rms noise in the spectra was about 1.2\% in terms of the continuum intensity.
The mosaics in Fig.~\ref{Fig:vttmaps} were each assembled
from ten slit-reconstructed images displaying the intensity at different
wavelengths: \mbox{H$\alpha+0.5$~\AA}, H$\alpha$ line core,
\mbox{H$\alpha-0.5$~\AA}, and \ion{Na}{i} D$_2$ line core. 

In addition to the spectroscopic observations, full-disk images of the
Chromospheric Telescope \citep[ChroTel,][]{bethge11} mounted on the flat roof of
the VTT building were used in this study. ChroTel acquires images at several
chromospheric wavelengths (H$\alpha$ $\lambda$6563~\AA, \ion{He}{i}
$\lambda$10830~\AA, and \ion{Ca}{ii}\,K $\lambda$3933~\AA) with a cadence of
3~min. Furthermore, full-disk magnetograms of the Helioseismic and Magnetic
Imager \citep[HMI,][]{HMI} on board the Solar Dynamics Observatory
\citep[SDO,][]{SDO} were used to interpret the spectra in the context of the
photospheric magnetic field.

Standard image reduction procedures were carried out for Echelle spectra and
ChroTel images including dark and flat-field corrections. The Echelle spectra 
was normalized to the continuum using the Fourier Transform Spectrometer (FTS) spectrum from the
Kitt Peak National Observatory \citep{neckel1984}. In addition, ChroTel images
were corrected for limb darkening. The ChroTel H$\alpha$ full-disk image is
presented in Fig.~\ref{Fig:ChroHMI} along with an HMI magnetogram. The extremely
long filament clearly followed the magnetic polarity inversion line (PIL), which
is faintly seen in the magnetogram. A more detailed magnetogram underlying the
filament is shown in Fig.~\ref{Fig:chrotelmaps}.


\section{Data analysis}\label{Sect:Dataanalysis}

High-resolution spectroscopic observations of giant filaments are rare
considering the time needed to scan a structure with a size comparable to the
solar radius and the excellent seeing conditions required for a good performance
of the adaptic optics system. In the following, we will describe the morphology of this
filament and derive the Doppler velocities from H$\alpha$ and \ion{Na}{i} D$_2$ line
shifts. Furthermore, we will focus on the H$\alpha$ line and use CM
inversions to infer some of the filament's plasma properties.


\subsection{Morphological description}\label{Sect:morphdescription}

What makes this filament special is its huge linear dimension. The H$\alpha$ filtergram 
in Fig.~\ref{Fig:vttmaps} shows that the
filament has a linear extension of $\sim$817\arcsec, which corresponds to about 658~Mm along a great circle
on the solar surface. In the following, positions are given in heliocentric coordinates
as provided on the axes of Fig.~\ref{Fig:vttmaps}. Close to its eastern end the
filament has a gap between $300$\arcsec~E and $250$\arcsec~E. The gap is even
larger in the \ion{He}{i} image depicted in Fig.~\ref{Fig:chrotelmaps}, and it 
is located where the filament is broadest. 
Multi-wavelength time-lapse movies
from the Atmospheric Imaging Assembly \citep[AIA,][]{AIA} show  at this location
a filament eruption on 2011 November~14, which was associated with a
non-geoeffective CME. Thin threads, visible both in
H$\alpha$ and \ion{He}{i}, fan out from both edges of the gap and indicate that
the filament returns to its original condition, when it was still a single
structure.

This filament exhibits two structural components: the spine and
barbs. The slit-reconstructed H$\alpha$
line core image in Fig.~\ref{Fig:vttmaps} clearly shows the spine of the
filament between $150$\arcsec~E and $125$\arcsec~W as well as several barbs,
i.e., threads that protrude from the main axis (see e.g., labels `c' and `d').
The same structure is also visible in the \ion{He}{i} line core image in
Fig.~\ref{Fig:chrotelmaps}. By contrast, the filament is barely seen in the
\ion{Ca}{ii}\,K line core image in Fig.~\ref{Fig:chrotelmaps}, although some
absorption is detected along the spine. Surprisingly, there is \ion{Ca}{ii}\,K
absorption in the gap between both filament portions supporting the notion that
plasma is again filling up the filament channel.

HMI magnetograms were used to obtain information about the photospheric magnetic
field below the filament. Quiescent filaments are found outside of active
regions. Hence, the magnetic signal below and in the close surroundings of the
filament is very weak. For this reason, a ``deep'' magnetogram was composed,
which is presented in Fig.~\ref{Fig:chrotelmaps}. All 45-second-cadence
magnetograms between 11:30 and 13:20~UT were corrected for differential rotation
and then averaged. Assuming that field lines in the quiet Sun are perpendicular
to the surface, the magnetic field was corrected for line-of-sight (LOS)
projection effects by dividing the magnetogram by $\mu = \cos \theta$, where
$\theta$ is the heliocentric angle. Clipping the magnetogram between $\pm 30$~G
reveals that the filament tightly follows the PIL. The polarities begin to mix
at the PIL at the left end of the bigger portion of the filament
($\sim$250\arcsec~E). A small bipolar emerging flux region (EFR) is present at
around 225\arcsec~N and 150\arcsec~W, which is close to the right end of the
filament (see rectangle in Fig.~\ref{Fig:vttmaps}). The EFR appears bright in the H$\alpha$ mosaic. 
However, this does not imply that the H$\alpha$ line 
was in emission, it only means that the line core is substantially
shallower, and also the line wing is enhanced, compared to line profiles belonging to the quiet Sun. In comparison,
the \ion{He}{i} line core image in Fig.~\ref{Fig:chrotelmaps} reveals strong
absorption at the same location.

\begin{figure}[t]
\includegraphics[width=\columnwidth]{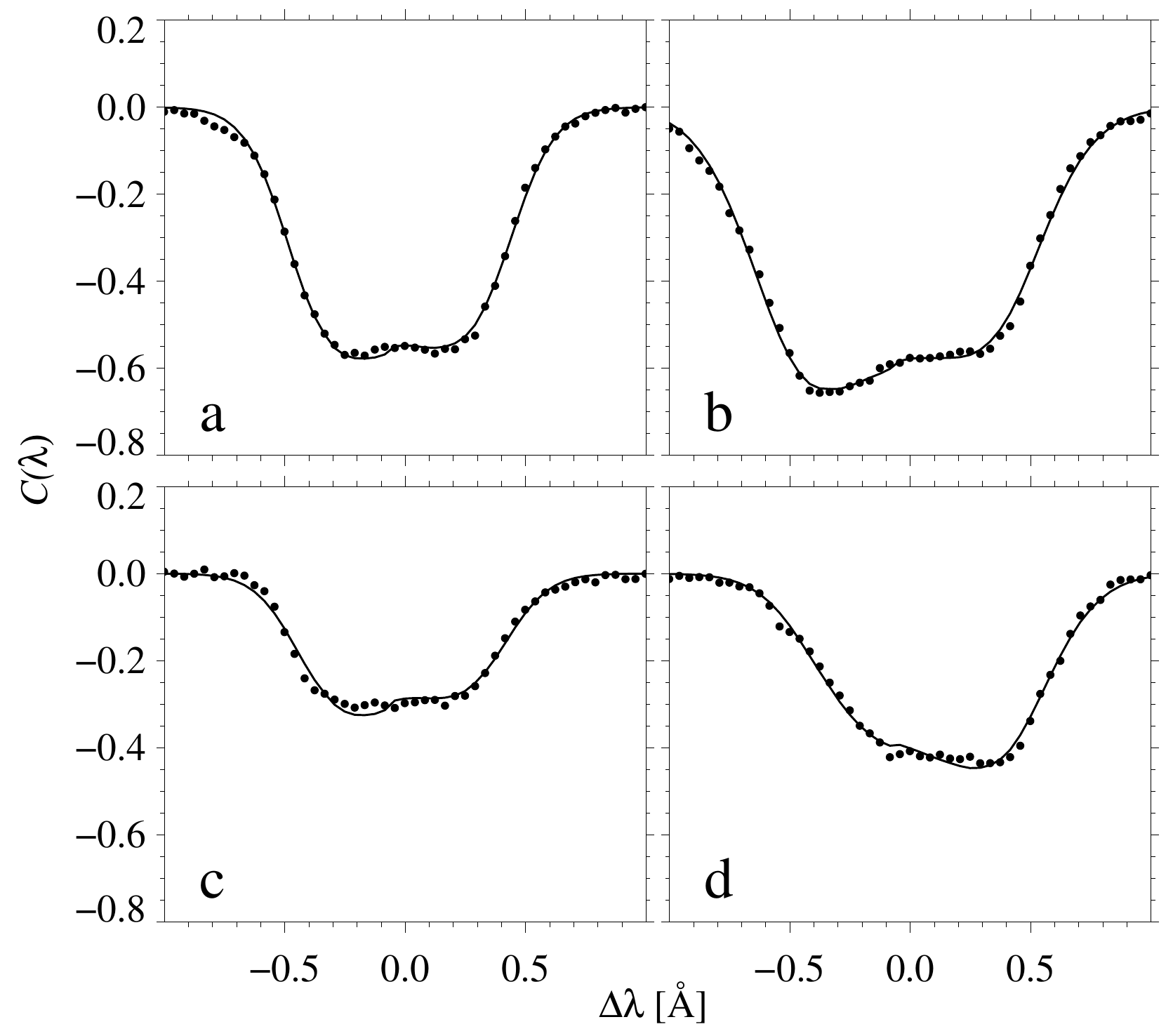}
\caption{Four selected H$\alpha$ contrast profiles. The dots represent
    the ``observed'' contrast profiles while the solid line shows the best fit from the CM inversion. 
    Panels `a' and `b' belong to very opaque regions of the filament seen in H$\alpha$, whereas
    panels `c' and `d' belong to the barbs of the filament. The exact locations are indicated in 
Fig.~\ref{Fig:vttmaps}.}
\label{Fig:cprofiles}
\end{figure}

The spine of the filament is visible in the off-band slit-reconstructed
images at \mbox{H$\alpha \pm 0.5$~\AA} depicted in Fig.~\ref{Fig:vttmaps}.
Interestingly, the spine is seen in both off-band images. Thus, the H$\alpha$
line profile is very broad in this area of the filament. This distinct
characteristic is also easily recognizable in contrast profiles belonging to the
spine, e.g., the one labeled `b' in Fig.~\ref{Fig:cprofiles}. This contrast
profile stands out because of its broad and deep shape. Furthermore, the
H$\alpha$ line-wing mosaics reveal information about the Doppler line shifts.
The barb labeled `d' appears prominently in the \mbox{H$\alpha + 0.5$~\AA} image,
whereas it is very faint in the \mbox{H$\alpha - 0.5$~\AA} image.
Consequently, the spectra belonging to this barb are mainly redshifted. 
Nevertheless, inside the barb, adjacent dark features, which are either seen in the \mbox{H$\alpha + 
0.5$~\AA} or in the \mbox{H$\alpha - 0.5$~\AA} images, 
are consistent with counter-streaming flows as already reported by \citet{diercke2014} using SDO/AIA 
data of the same filament.
Many but not all barbs show a similar behavior. For example, the barb at 
$175$\arcsec~E is darker in the \mbox{H$\alpha - 0.5$~\AA} image than in the
\mbox{H$\alpha + 0.5$~\AA} image. Thus, this area of the filament is
blueshifted. These findings will be compared with the inferred velocities from the CM in
Sect.~\ref{Sect:cloudmodelresults}.

Brightenings accompany the filament on both sides in the \ion{Na}{i} D$_2$ mosaic.
A comparison with the HMI magnetograms reveals that at least the largest 
conglomerates of brightenings coincide with small-scale magnetic fields. 
Hence, these brightenings can be used as a tracer of the magnetic field.  
In the \ion{Na}{i} D$_2$ line core images, neither signs of the spine nor of the barbs 
are apparent.


\subsection{Line-of-sight H$\alpha$ and Na\,{\scriptsize I} D$_2$
    velocities}\label{Sect:LOS}

Doppler shifts of the H$\alpha$ and \ion{Na}{i} D$_2$ line cores were derived by 
using a least-squares parabola fit. On average, the selected width of the broad H$\alpha$ line core for the parabola
fit spanned $\sim 750$~m\AA. A smoothing over 8 pixels in the spectral dimension was performed twice.
Because the H$\alpha$ line originates in the
chromosphere, it is not affected by the convective blueshift. Thus, the
assumption that up- and down-flows are balanced in the quiet-Sun chromosphere is
justified to establish a reference for the inferred Doppler velocities. 

The \ion{Na}{i} D$_2$ line contains a noticeable line-blend between the core and the
blue wing. Therefore, we used a narrower region around the line core ($\sim $145~m\AA) to derive the
Doppler velocities. Previous to the parabola fit we slightly
smoothed the spectra. For this purpose, a finer method than the one applied to the H$\alpha$ line, 
consisting of a convolution with a normalized Gaussian with a FWHM of 17~m\AA, was carried out. The \ion{Na}{i} D$_2$ 
reference
line-center position is also the average position of all quiet-Sun profiles within one map. Since the filament covers 
such a large area on the Sun, global velocity gradients have been removed by subtracting a linear fit to each of 
the ten velocity maps. This effect was negligible for the H$\alpha$ line which in general showed larger Doppler shifts. 
Hence, the subtraction was only done for the velocities inferred from the \ion{Na}{i} D$_2$ line. 

\subsection{Cloud model inversions}\label{Sect:cloudmodelinv}

In his seminal work, \citet{beckers64} introduced the \textit{Cloud Model}
to analyze spicules observed in H$\alpha$. This approach can also be
applied to H$\alpha$ spectra of filaments or any type of feature exhibiting clouds of absorbing
plasma in the chromosphere \citep[see also][for a 
review on CM inversion techniques]{tzio07}. As in spicules, there is a
significant difference between the contrast of H$\alpha$ intensity profiles
inside and in the surroundings of filaments. This property can be used to solve
the radiative transfer equation in a simplified manner. The 
three CM assumptions are: (1) the physical parameters are constant in height, (2) the
source function $S$ does not depend on the wavelength, and (3) the line
absorption coefficient has a Gaussian behavior. Under these circumstances, the
radiative transfer equation can be written as
\begin{equation} \label{eq:rt}
C(\lambda) \equiv \frac{I(\lambda) - I_0(\lambda)}{I_0(\lambda)} = 
    \left(\frac{S}{I_0(\lambda)} - 1 \right)
    \Big[1 - \exp\Big(-\tau(\lambda)\Big)\Big], 
\end{equation}
where $C(\lambda)$ is the contrast profile, $I(\lambda)$ is an observed spectral
profile, $I_0(\lambda)$ is the mean quiet-Sun spectral profile, $S$ the average
source function of the cloud, and $\tau(\lambda)$ is defined as
\begin{equation} \label{eq:tau}
\tau(\lambda) = \tau_0 \exp{\left(-\left( \frac{\lambda - \lambda_\mathrm{c}}{\Delta
    \lambda_\mathrm{D}}\right)^2\right)} ~.
\end{equation}
The optical thickness of the cloud is constant and denoted by $\tau_0$. The parameters inside
the brackets in Eq.~(\ref{eq:tau}) are the central wavelength $\lambda_\mathrm{c}$ and
the Doppler width $\Delta \lambda_\mathrm{D}$ of the line. The LOS velocity
of the absorbing cloud material can be derived as
\begin{equation} \label{eq:Vlos}
 v = \frac{\lambda_\mathrm{c} - \lambda_0}{\lambda_0 } c,
\end{equation}
where $\lambda_0$ is the central wavelength of the H$\alpha$ line and $c$ the
speed of light.

The contrast profiles $C(\lambda)$ can be calculated for each pixel using the
mean quiet-Sun profile $I_0(\lambda)$ and the observed intensity profile
$I(\lambda)$. For this purpose, a smaller spectral range was selected, which widely covered the H$\alpha$ line. 
The spectral range is indicated by the shaded area in Fig. \ref{Fig:Haprofile}.
The observed profiles $I(\lambda)$ were convolved with a
normalized Gaussian ($\mathrm{FWHM} = 9.8$~m\AA) to
slightly smooth the profiles. For the quiet-Sun profile, only pixels with
low-contrast H$\alpha$ structures were selected. These spectra were aligned with
respect to an arbitrary spectrum to avoid line broadening and then the mean
quiet-Sun profile was computed. Four examples of contrast profiles are
presented in Fig.~\ref{Fig:cprofiles}. Their exact location 
is indicated by labels `a'\,--\,`d' in Fig.~\ref{Fig:vttmaps}. The number of
spectral points was reduced to 79 employing a 10-pixel binning (dots in
Fig.~\ref{Fig:cprofiles}). The trimmed-down number of spectral points is still
sufficient for reliable CM fits.

\begin{table}
\caption{Inferred CM parameters from the contrast profiles shown in Fig.~\ref{Fig:cprofiles}.}
\label{Tab:contrastprofs}      
\centering                                      
\begin{tabular}{c|c c c c}          
\hline\hline                        
$C(\lambda)$  & $\tau_0$  & $v$             & $\Delta \lambda_\mathrm{D}$ & $S$ \rule[1mm]{0mm}{2mm}\\    
              &           &  [km\,s$^{-1}$] &         [\AA]               & \\    
\hline 
      & \multicolumn{4}{c}{Filament}   \rule[1mm]{0mm}{2.5mm}  \\
\hline                       
    a & 2.97 & $-1.06$ & 0.34 & 0.07  \rule[1mm]{0mm}{2.6mm} \\     
    b & 2.77 & $-3.11$ & 0.45 & 0.06\\
    c & 1.46 & $-0.84$ & 0.33 & 0.10 \\
    d & 1.42 & $+4.41$ & 0.40 & 0.07 \\
\hline                              
\end{tabular}
\end{table}

The aim of CM inversions is to fit each contrast profile $C(\lambda)$
by modifying the four parameters of Eq.~(\ref{eq:rt}): $\tau_0$, $v$, $\Delta
\lambda_\mathrm{D}$, and $S$. To increase the efficiency and to improve the
convergence of the fitting algorithm, it is necessary to provide a reasonable
initial estimate of the input parameters. A data base of 50000 contrast profiles
was created subject to the following limits for the randomly chosen input
parameters: $\tau_0 \in [0, 3]$, $v \in [-38, 38]$~km~s$^{-1}$, $\Delta
\lambda_\mathrm{D} \in [0.08, 0.70]$~\AA, and $S \in [0, 0.4]$. These limits
were based on former cloud-model studies regarding filaments
\citep[e.g.,][]{alissandrakis90, chae06}. Each observed contrast profile is
cross-checked against the templates in the data base, and the parameters yielding
the lowest $\chi^2$-value are selected for Levenberg-Marquardt least-squares
minimization \citep[][]{more1977}. The solid lines in Fig.~\ref{Fig:cprofiles}
represent the best fits to the respective H$\alpha$ contrast profiles.
The fit parameters are given in Table~\ref{Tab:contrastprofs}.


\section{Results}\label{Sect:Results}


\begin{SCfigure*}[0.68][t]
\includegraphics[width=0.68\textwidth]{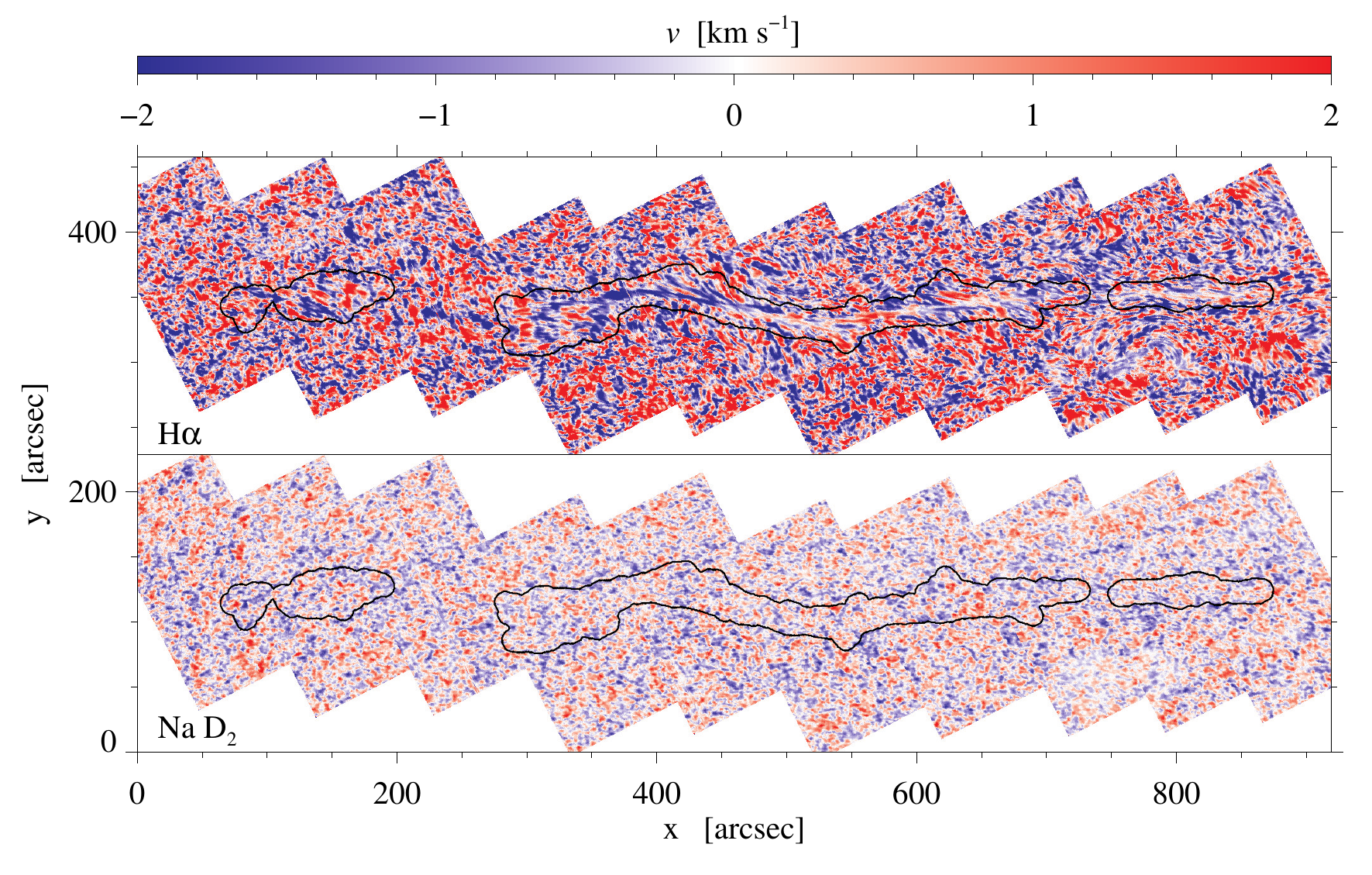}
\hspace*{3mm}\vspace{0mm}
\caption{Mosaic of H$\alpha$ (\textit{top}) and \ion{Na}{i} D$_2$ (\textit{bottom})
    Doppler velocities corresponding to the slit-reconstructed images in 
    Fig.~\ref{Fig:vttmaps}. The velocities were inferred from the line core fits.  
    The mosaics were rotated to place the filament in a
    horizontal position. The superimposed black contours mark the outer
    boundaries of the filament as seen in H$\alpha$ line core images. Doppler
    velocities were clipped between $\pm$2~km~s$^{-1}$.}
\label{Fig:vmaps}%
\end{SCfigure*}

\begin{table}
\caption{H$\alpha$ and \ion{Na}{i} D$_2$ average LOS velocities ($\overline{v}$), 
standard deviation ($\sigma$) and number of points (\#) used for the statistics. 
Maps 5, 6, and 7 in Fig. \ref{Fig:vttmaps} where used for the spine statistics. 
The average quiet-Sun velocity was computed using maps 1--3 
(excluding pixels belonging to the filament and bright points seen in \ion{Na}{i} D$_2$).}            
\label{Tab:LOSvelocities}
\centering                       
\begin{tabular}{c|c c c c}       
\hline\hline                     
                           & \multirow{2}{*}{Line}      &  $\overline{v}$ & $\sigma$       & \# \\    
                           &                            & [km\,s$^{-1}$] & [km\,s$^{-1}$] & points \\
\hline                                  
\multirow{2}{*}{Filament}  & H$\alpha$          & $-0.225$ & 1.220 & \multirow{2}{*}{175322} \\ 
                           & \ion{Na}{i} D$_2$  & $-0.023$ & 0.606 &     \\
\hline
\multirow{2}{*}{Spine}     & H$\alpha$          & $-0.288$ & 1.133 & \multirow{2}{*}{71087}  \\
                           & \ion{Na}{i} D$_2$  & $+0.010$ & 0.588 &     \\
\hline
\multirow{2}{*}{Quiet Sun} & H$\alpha$          & $-0.070$ & 1.436 & \multirow{2}{*}{797224}  \\
                           & \ion{Na}{i} D$_2$  & $+0.042$ & 0.622 &     \\
\hline                                            
\end{tabular}
\end{table}

Velocity patterns and morphological changes characterize the evolution of (large-scale) filaments. 
Therefore, firstly we will present the LOS velocities based on spectral line profiles to retrieve the plasma 
properties for chromospheric features. In addition to that, we will show the physical parameters inferred 
from the CM inversions and compare them with previous works of smaller filaments. The coupling between the barbs of the 
filament and the underlying magnetic field is analyzed using HMI magnetograms. Finally, photospheric horizontal proper 
motions provide information of the flows below the filament.

\subsection{Line-of-sight velocities}
\begin{figure}[!t]
\resizebox{\hsize}{!}{\includegraphics{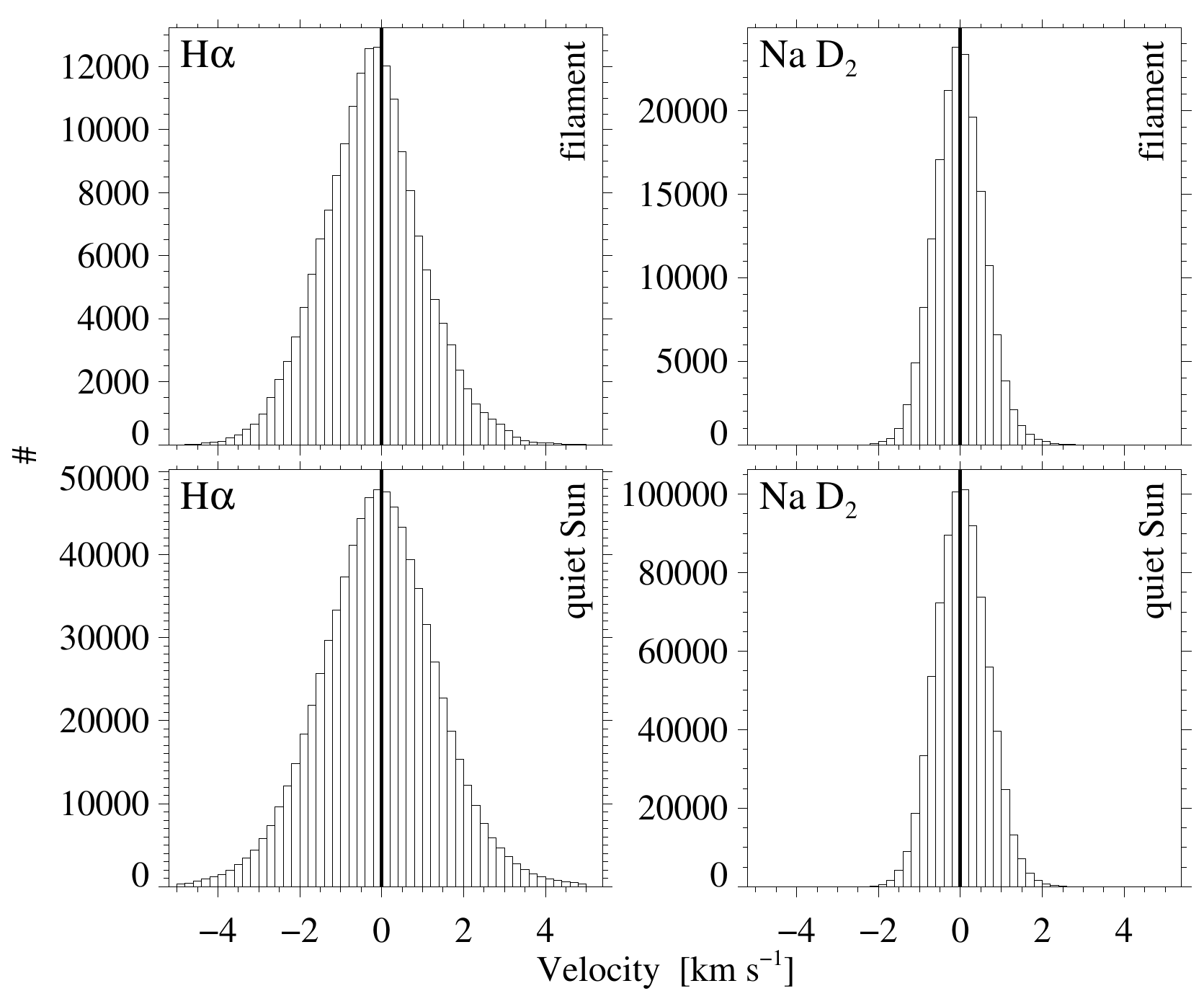}}
\caption{Frequency distributions of the Doppler velocities inferred from the line core fits of the H$\alpha$ (left) 
and \ion{Na}{i} D$_2$ (\textit{right}) lines in the filament (\textit{top}) and in the quiet Sun (\textit{bottom}). The 
bin size is 0.2\,km\,s$^{-1}$. }
\label{Fig:velhistograms}
\end{figure}

A map of the inferred LOS velocities performing line core fits is shown in Fig. \ref{Fig:vmaps}. The smooth black 
contour generously outlines the border of the filament. 
In the upper (lower) panel, the velocities retrieved from the H$\alpha$ (\ion{Na}{i} D$_2$) Doppler shifts are shown 
clipped between $\pm 2$~km\,s$^{-1}$. For H$\alpha$, as a general trend the spine of 
the filament exhibits blueshifts, i.e., upward flows. 
This is also confirmed when calculating the average Doppler shifts along all the spine (maps 5--7), as 
presented in Table \ref{Tab:LOSvelocities}. On average, the spine moves upwards at $-0.288$~km\,s$^{-1}$. 
The average LOS velocity of the whole filament is also blueshifted ($-0.225$~km\,s$^{-1}$), which can also be seen in 
the upper-left panel of the histogram in Fig. \ref{Fig:velhistograms}. The average quiet-Sun velocity frequency 
distribution was computed using maps 1--3 (excluding pixels belonging to dark structures of the 
filament and bright points seen in the \ion{Na}{i} D$_2$ line core image for both lines) and has roughly a Gaussian 
shape
(lower-left panel in Fig. \ref{Fig:velhistograms}). Its average value is 
slightly blueshifted yielding $-0.07$~km\,s$^{-1}$. Interestingly, the righthand gap of the
filament does not show significant changes in velocity. If one would remove the H$\alpha$ contours, it would
not be possible to distinguish the gap. On the contrary, this is not the case with the lefthand gap, which is much 
larger, because the velocities show a different pattern as compared to the spine.

The average LOS velocities inferred from the \ion{Na}{i} D$_2$ Doppler shifts inside the filament are one order of 
magnitude smaller  ($-0.023$~km\,s$^{-1}$) compared to the H$\alpha$ ones. The \ion{Na}{i} D$_2$ line seems not to be
greatly affected by the filament. Another evidence for this is that no alteration of the velocity pattern owing to the 
presence of the filament can be seen in Fig.~\ref{Fig:vmaps} (bottom panel). The velocity histograms on the righthand 
side of Fig.~\ref{Fig:velhistograms} reveal that the range of velocities is approximately a factor of 2 smaller than 
for H$\alpha$.


\subsection{Cloud model results}\label{Sect:cloudmodelresults}

A total of 175322 H$\alpha$ contrast profiles, which covered the giant filament region, 
were subjected to CM inversions as described in the previous section. After rejecting failed inversions, 175119 
were left for the statistical analysis. The present study is 
statistically more complete than previous works because of the large amount of inverted contrast profiles. 
The most common type of contrast profiles have typically an 
$\omegaup$-like shape (see Fig.~\ref{Fig:cprofiles}). 
When computing a correlation among all contrast profiles and 
considering that their correlation has to reach at least 95\%, up to 42.7\% show an 
$\omegaup$-like shape. The second most common type (37.1\%) is similar to the $\omegaup$-like shape 
but without the central hump. The $\omegaup$-like shape contrast profiles dominate, but are not
exclusively present, in the spine of the filament, i.e., in the second half of map 5 and along maps 6 and 7 
(see Fig. \ref{Fig:vttmaps}). Interestingly, barbs `c' and `d' are also dominated by the most common type of contrast
profiles. However, the barb-like structures seen in map 5 are more widely populated by the second 
most common type of contrast
profiles. The other parts of the filaments do not have a clear tendency towards a specific type of contrast-profile 
shapes. Variations of the H$\alpha$ profiles caused by unresolved counter-streaming flows might slightly modify 
some shapes of contrast profiles. In the sample, 
the depth of the contrast profiles can reach up to $C(\lambda) = -0.65$.

Figure~\ref{Fig:cprofiles} illustrates
that contrast profiles belonging to the darkest H$\alpha$ filament structures
(profiles `a' and `b') are deeper compared to the ones belonging to the threads
of the filament (profiles `c' and `d'). The inferred physical parameters for these profiles `a' and `b' are shown 
in Table \ref{Tab:contrastprofs}. The darker profiles `a' 
and `b' show consistently larger values of the optical thickness as compared to the threads, 2.97 and 2.77, 
respectively.

\begin{SCfigure*}[0.7][!t]
\includegraphics[width=0.7\textwidth]{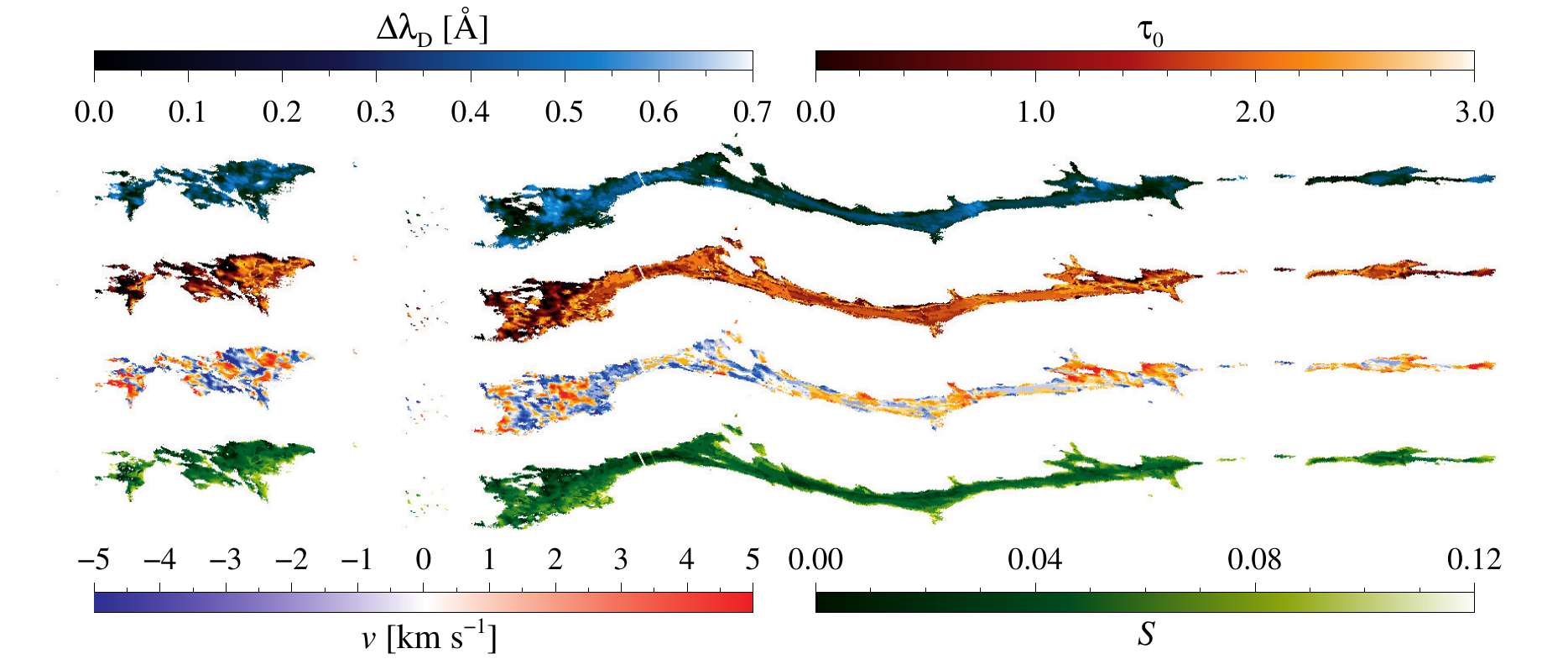}
\hspace*{3mm}\vspace{0mm}
\caption{From (\textit{top} to \textit{bottom}), maps of Doppler width $\Delta \lambda_\mathrm{D}$, optical thickness
    $\tau_0$, LOS velocity $v$, and source function $S$ of the filament inferred from the H$\alpha$ CM inversions.}
\label{Fig:cloudresults}
\end{SCfigure*}

The parameters retrieved from the CM inversions are presented in
Fig.~\ref{Fig:cloudresults}. The maps reveal that the parameters are not
uniformly distributed along the filament. There is a good correlation
between the LOS velocities and the \mbox{H$\alpha \pm 0.5$~\AA} intensity maps
displayed in Fig.~\ref{Fig:vttmaps}. 
For instance, barb `d' appears dark in the \mbox{H$\alpha + 0.5$~\AA} image. As a consequence, this thread
appears mainly redshifted in the inferred velocity map. 
Indeed, the inferred CM value yields 4.41~km\,s$^{-1}$ (see Table~\ref{Tab:contrastprofs}). Hence, material is flowing 
downwards, although some minority upward trend is also seen. Yet not all barbs show the same behavior. The group of 
barbs left to barb `c' (map 5 in Fig. 
\ref{Fig:vttmaps}) is more dominated by blueshifts. Therefore, one cannot conclude from this giant filament whether 
barbs systemically contribute to mass supply or loss of filaments. Counter-streaming flows inside barbs are a good 
candidate to explain this behavior.

The \mbox{H$\alpha \pm
0.5$~\AA} intensities reveal information about the line shifts. Patches with
anti-parallel directions of the LOS velocities occur in close proximity. A possible explanation
are waves that move along the filament, similar to the vertical oscillatory
motions found by \citet[][]{okamoto07}. However, in the absence of time-series
data this hypothesis cannot be confirmed. 

\begin{figure}[t]
\resizebox{\hsize}{!}{\includegraphics{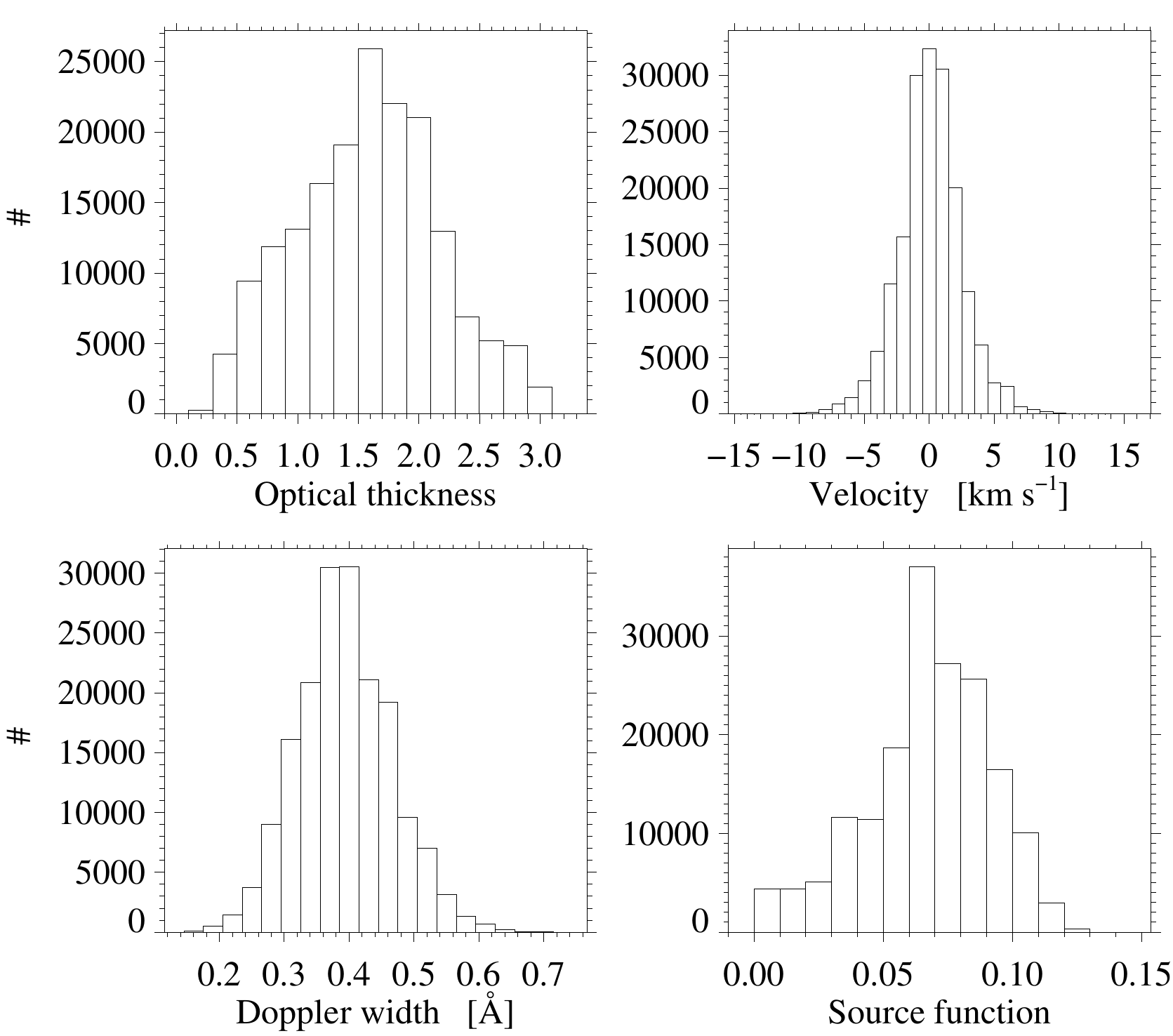}}
\caption{Frequency distributions of the four CM parameters: optical
    thickness $\tau_0$, LOS velocity $v$, Doppler width $\Delta
    \lambda_\mathrm{D}$, and source function $S$ (\textit{top-left} to
    \textit{bottom-right}).}
\label{Fig:cloudhistograms}
\end{figure}

Histograms of optical thickness, LOS velocity, Doppler width, and source function are shown in Fig.
\ref{Fig:cloudhistograms}. 
The histograms were computed independently for each scan and then added to present statistics over
the whole filament. The fact that there was always an overlap between one map and the 
following while scanning the filament, introduces data points which were counted twice. However,
although the overlapped areas were co-spatial, they were not scanned at the same time. Therefore, the
information encoded in the contrast profiles is new. 

\begin{table}
\caption{Statistics of the H$\alpha$ CM parameters. The total number of inverted profiles was $n = 175119$.}        
\label{Tab:cloudstats}      
\centering                                      
\begin{tabular}{c c c c c}          
\hline\hline                        
Parameter & Mean   & Std dev ($\sigma$)  & Skewness & Kurtosis \rule[1mm]{0mm}{2.6mm}  \\    
\hline                                   
$\tau_0$   &  1.59  & 0.59      & $+0.03$     & $-0.52$  \rule[1mm]{0mm}{2.6mm}   \\      
$\Delta \lambda_\mathrm{D}$ [\AA] & 0.39& 0.07&$+0.20$& $+0.19$    \\    
$v$ [km\,s$^{-1}$]& 0.07  & 2.46      & $-0.05$  & $+1.32$       \\ 
$S$       & 0.07  & 0.02      & $-0.49$  & $+0.03$       \\
\hline                                             
\end{tabular}
\end{table}

The histograms together with their statistics in Table \ref{Tab:cloudstats} reveal some interesting facts which will be
compared to previous works from \citet[][]{schmieder03} and \citet[][]{chae06}. Both studies included CM inversions
of H$\alpha$ contrast profiles in a smaller quiescent filament. 
At a first glance, the histograms of optical thickness and source function resemble a triangular shape, whereas the
velocity and Doppler width have a Gaussian shape. The optical thickness has a mean value of 1.59, which is rather 
large compared to 0.66 and $0.83\pm0.47$ retrieved from the cloud model inversions 
with a constant source function of \citet[][]{schmieder03} and \citet[][]{chae06}, 
respectively. The LOS velocity histogram is clearly centered at rest ($\sim 0.07$~km\,s$^{-1}$), with a quite large
dispersion of $\sigma = 2.46$~km\,s$^{-1}$. 
The dominant Doppler width values are in the range 0.36 and 0.42~\AA, on average 0.39~\AA, with a positive 
skewness. 
This range closely agrees with the two filaments studied by
\citet[][]{schmieder03} and \citet[][]{chae06}. The source function peaks at around 0.07, with negative skewness, 
which lies in between the values found by \citet[][]{schmieder03} and \citet[][]{chae06},i.e.,  
0.05 and 0.10, respectively. 

Our results do not necessarily have to match the outcome of other works. On the one hand, the filament 
under study is peculiar because it has extraordinary linear dimensions and might also be in a different stage of its 
lifetime compared to the other two filaments. On the other hand, stray-light or different instrument profiles can affect 
the CM inversions, since they are sensitive to the line depth and shape. In order to test these
effects, we performed two additional CM inversions of map~6. In the
first inversion we added 5\% stray-light to the observed profiles and in
the second one we convolved the observed profiles with a Gaussian (with
a FWHM of 50~m\AA) simulating a different instrument, e.g., a Fabry-P\'erot interferometer. We conclude that
adding stray-light only effects the source function increasing its value by the same amount as the added 
stray-light. All other CM parameters remain almost the same. 
When convolving with a Gaussian, we found an increase of $\sim
3$\% of the mean optical thickness compared to the original inversions.
The velocity trends were conserved. In conclusion, the inferred optical
thickness can indeed be effected by different instrument profiles.
However, we expect that within the above limitations the retrieved mean value
of the optical thickness can be compared among different instruments. In addition, the selection of a proper quiet-Sun 
profile is crucial \citep[][]{bostanci10},
because it also effects the outcome of the CM inversions and can be of the same order
of magnitude as additive stray-light or different instrument profiles.


\subsection{Reconstitution of the filament within the filament channel}
On the lefthand side of the filament, 
a large gap is seen in the H$\alpha$ and \ion{He}{i} $\lambda$10830~\AA\ images. The gap appeared after part of 
the filament erupted the day before. Remarkably, fine H$\alpha$ threads which appeared 
at the edges of the gap were detected. These threads seemed to indicate that the filament was filling up the gap again 
with plasma. H$\alpha$ filtergrams of the Kanzelh\"ohe Observatory from November~15 also showed 
dynamic plasma features appearing in the gap, mainly close to the edges. ChroTel filtergrams of the \ion{Ca}{ii}~K 
line confirmed this hypothesis as absorption of this line was detected inside the gap. We therefore 
suggest that the filament was in the process of reestablishing its initial configuration.


\subsection{Coupling between barbs and the photospheric magnetic field}
Is there evidence for a connection between the barbs of the filament and the underlying magnetograms from HMI? 
In the past, it has been reported that barbs are rooted in patches of minority polarity among the dominant polarity on 
each side of the filament \citep[][]{martin98}. In a later study, \citet[][]{chae05} found evidence that barbs end at 
minor polarity inversion lines. Similarly, \citet[][]{lopezariste06} showed that the endpoints of barbs are next to 
parasitic polarity features.


\begin{figure}[t]
\resizebox{\hsize}{!}{\includegraphics{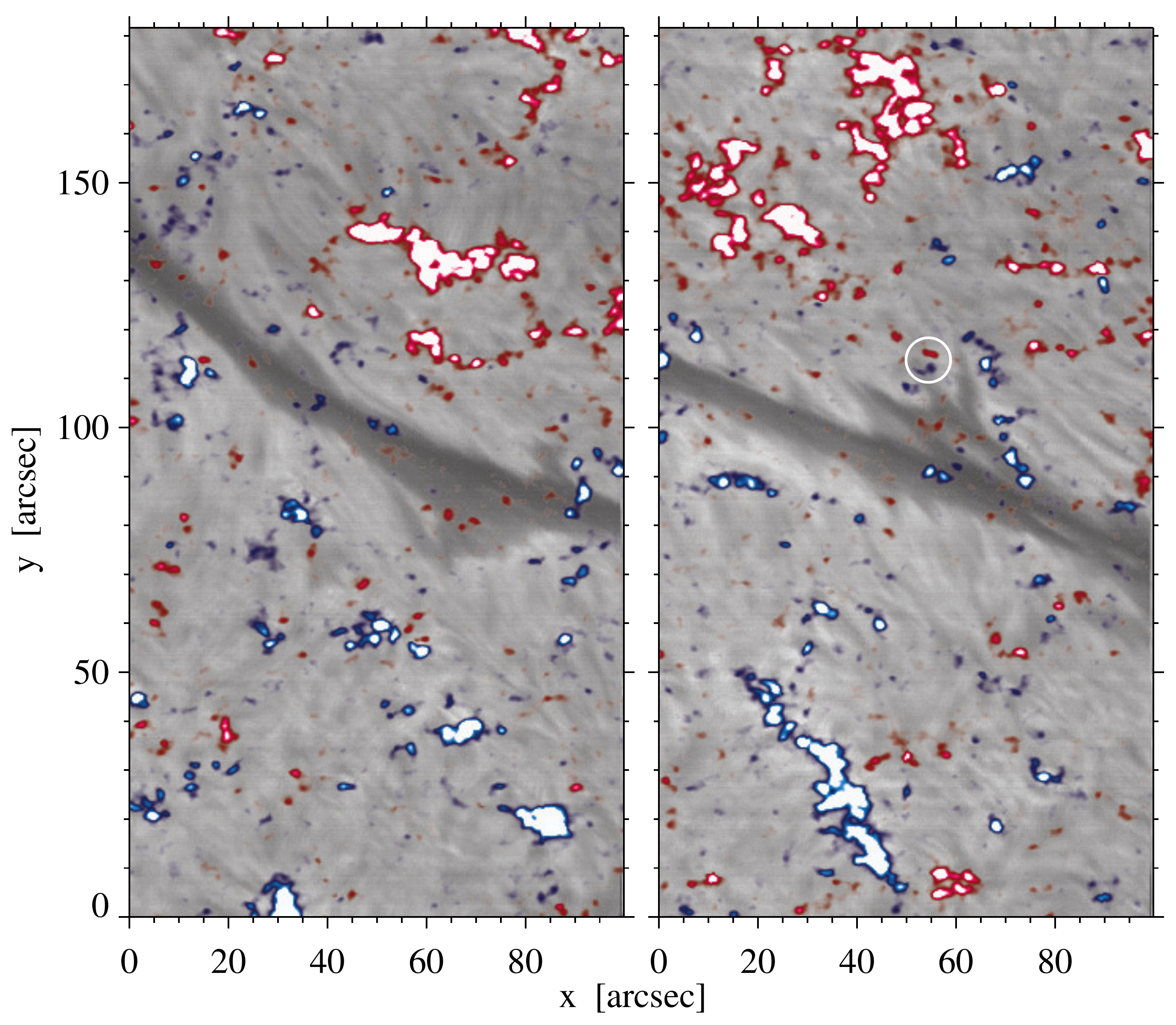}}
\caption{Overview of the barbs from the filament seen in maps 6 and 7 of the slit-reconstructed H$\alpha$ line core 
data from Fig. \ref{Fig:vttmaps}. Overlying is the HMI magnetogram clipped at 25~G. The color scale goes from blue 
(negative polarity) to red (positive polarity) and the white color represents saturated values.}
\label{Fig:HaHMIoverlap}
\end{figure}


To address this question, a thorough alignment among the slit-reconstructed 
H$\alpha$ line core images and an averaged deep magnetogram was performed. 
The magnetic field below the filament was weak. 
To increase the signal several magnetograms were corrected and averaged as described in Sect. 
\ref{Sect:Dataanalysis}. The magnetograms were chosen to match, in time and space, the 10-min scan of the VTT (a 
total amount of 14 images). However, to scrutinize very small magnetic fields close to the ends of barbs we 
added 13 min before and after the VTT scan, increasing the number of magnetograms to 48 ($\sim$36~min). We noticed 
that the magnetic structures did not wash out or cancel each other when increasing the number of images from 14 to 48. 

There is a good correlation between bright features of the slit-reconstructed line core \ion{Na}{i} D$_2$ images 
and magnetic structures in the magnetograms. Therefore, the \ion{Na}{i} D$_2$ images were used as an intermediate step 
for the alignment. 
The H$\alpha$ and \ion{Na}{i} D$_2$ images are well aligned and only differ on a subpixel scale. 
Figure~\ref{Fig:HaHMIoverlap}
shows the barbs as seen in the H$\alpha$ line core corresponding to maps 6 and 7. Superimposed is the 
corresponding averaged HMI magnetogram with a blue (negative polarity) to red (positive polarity) color scale clipped 
at 25~G. The faintest magnetic structures have 2~G while stronger fields ($>25$~G) appear saturated with white color. 
The filament is tightly confined by the stronger polarities, which lie on each side of the spine. Mixed polarities can 
be seen along the whole filament and even further away. Interestingly, as reported by the aforementioned 
authors and more recently by \citet[][]{joshi13}, 
these mixed polarities also appear at some ends of barbs, e.g., inside the white circle of the righthand panel of Fig. 
\ref{Fig:HaHMIoverlap}. However, this does not happen everywhere and can therefore not be generalized to all barbs. We 
can at least confirm that this property is also present in extremely large filaments. In this particular case, the 
magnetogram shows fields up to $\sim$14~G inside the white circle.


\subsection{Horizontal proper motions}

 \begin{figure*}
\includegraphics[width=\textwidth]{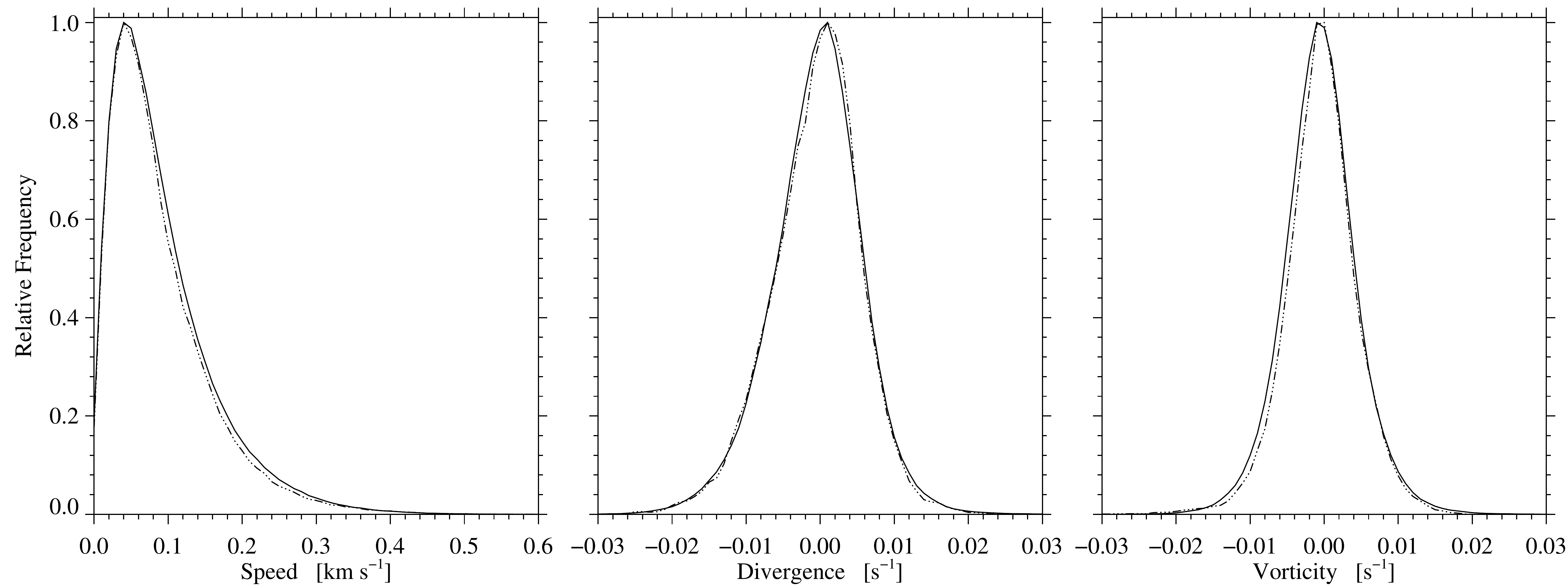}
\caption{Frequency distributions for speed (\textit{left}), divergence 
(\textit{middle}), and vorticity (\textit{right}) of horizontal proper motions 
computed by applying DAVE to HMI magnetograms. The number of points that went 
into the computation of the frequency distributions for the quiet Sun 
(\textit{solid line}) and the filament (\textit{dash dotted}) are $n_\mathrm{qs} 
= 2772670$ and  $n_\mathrm{fil} = 149007$, respectively.}
\label{Fig:DAVE}
\end{figure*}

We computed the horizontal proper motions using LOS magnetograms from 
HMI. We took one-hour time-series of magnetograms centered on the time 
12:22~UT. We applied the Differential Affine Velocity Estimator 
\citep[DAVE;][]{Schuck2005, Schuck2006} to geometrical foreshortening corrected 
magnetograms to retrieve flow fields. After this 
correction the FOV of the magnetograms was 649 $\times$ 466~Mm. The code 
DAVE assumes an affine velocity profile, and the underling physics is based on 
the magnetic induction equation. The horizontal flows are estimated using 
temporal and spatial derivatives of the magnetic field.
We used the \citet{Scharr2007} kernel for the spatial derivatives, a five-point stencil for the temporal derivative, 
and a sampling window of 11~pixels (3.5~Mm) \citep{Balthasar2014} in the computation of the velocity by DAVE.
The computation of the 
temporal derivatives utilized a sliding average of 16 magnetograms corresponding to 12 
min. In the end, the velocities were averaged for an hour.

In the DAVE velocity map we observed regular, diverging supergranulation cells as 
previously seen by \citet{Magara1999}, \citet{Rondi2007}, and 
\citet{Schmieder2014} in velocity maps computed using photospheric continuum 
images. The mean quiet-Sun velocity of 0.096$\pm$0.069~km~s$^{-1}$ is within the range given 
by \citet{diercke2014}. In general, the divergence and vorticity based on 
flux-transport velocities are an order of magnitude larger than those derived 
from LCT measurements \citep{Verma2012a}. The global properties of the quiet Sun 
and the photosphere below the observed filament are virtually the same. The 
respective frequency distributions for flow speed, divergence, and vorticity in 
Fig.~\ref{Fig:DAVE} are almost identical. The velocity distribution can be 
approximated with a log-normal probability density function (PDF), and the 
divergence and vorticity distribution obey both Gaussian PDFs (centered at zero 
and with FWHM of 0.012 and 0.009~s$^{-1}$, respectively). Any differences between quiet-Sun 
and filament distributions are most likely selection effects. For example, 
picking randomly an area of similar size as the filament region leads to 
comparable deviations of the frequency distributions. Only the EFR exhibits 
horizontal proper motions on larger scales, which are significantly different 
from the quiet-Sun flow field, i.e., they possess higher flow speeds and contain 
extended regions of positive divergence. 

The likeness of photospheric horizontal flow fields for quiet Sun and filaments 
hampers detecting any peculiar flows in and around the giant filament by visual 
inspection \citep[cf.,][]{Rondi2007}. Only by close examination, we identified 
several flow kernels with a size of 5--8~Mm located at the ends of the 
filament with speeds of up to 0.30--0.45~km~s$^{-1}$. These kernels consist of 
closely spaced sinks and sources, whereby sinks are predominantly confined to 
the photospheric region below the filament and sources to the exterior. Such 
a configuration contributes to the enhancement of the magnetic field at the ends 
of the filament. In particular, this might be an indication that the filament is 
in process to fill the gap between the eastern and western part, which was 
previously expelled during a CME. This finding is in agreement with 
\citet{Schmieder2014}, who also find converging motions at the edges of the 
filament channel especially near footpoints of the filament.


\section{Discussion and conclusions}\label{Sect:Conclusions}
The excellent seeing conditions on 2011 November~15 lasted the entire morning
until well after local noon, which was a fortuitous occasion allowing us to
capture this data set. The adaptive optics reliably locked on granulation and ten consecutive scans with 
high-resolution Echelle spectroscopy covered the giant filament which had linear dimensions of $\sim$817\arcsec\ (658 Mm 
along a great circle).  
 
We performed CM inversions using the H$\alpha$ line and retrieved valuable information 
about the physical parameters inside the filament. A total
amount of 175119 H$\alpha$ contrast profiles, covering the whole filament, were inverted with very reasonable fits. 
\citet[][]{chae06} showed a 
contrast profile located inside the spine of the filament with almost the same shape than ours. The depth of our 
contrast profiles reached up to $C(\lambda) = -0.65$. Deeper contrast profiles were related to darker H$\alpha$ 
structures of the filament and hence showed larger values of the optical thickness. While the average values inferred 
from the CM inversions for the LOS velocity, Doppler width, and source function were similar or lay in between the 
values found by \citet[][]{schmieder03} and \citet[][]{chae06}, the mean optical thickness was much larger in our data 
set ($\overline{\tau_0} = 1.59$).
Caution should be taken when comparing the 
physical parameters inferred from CM inversions between different instruments. We have shown that stray-light heavily 
affects the results of the source function. In addition, simulating our profiles as observed by a different instrument 
with less spectral resolution yielded variations of the mean optical thickness. Nevertheless, these variations on 
average were small ($< 3$\%).
The analyzed filament is of particular interest because of its large 
horizontal extension. From a speculative point of view, huge filaments might as well have an increased optical 
thickness. We have seen that the filament under study has an average optical thickness which is 
larger compared to the average value provided by \citet[][]{schmieder03} and \citet[][]{chae06} of smaller filaments. 
Optical thick areas were mainly, but 
not exclusively, found along the spine and even in barbs (e.g., see barb `d' in Fig.~\ref{Fig:cloudresults}).

\begin{figure}[t]
\includegraphics[width=\columnwidth]{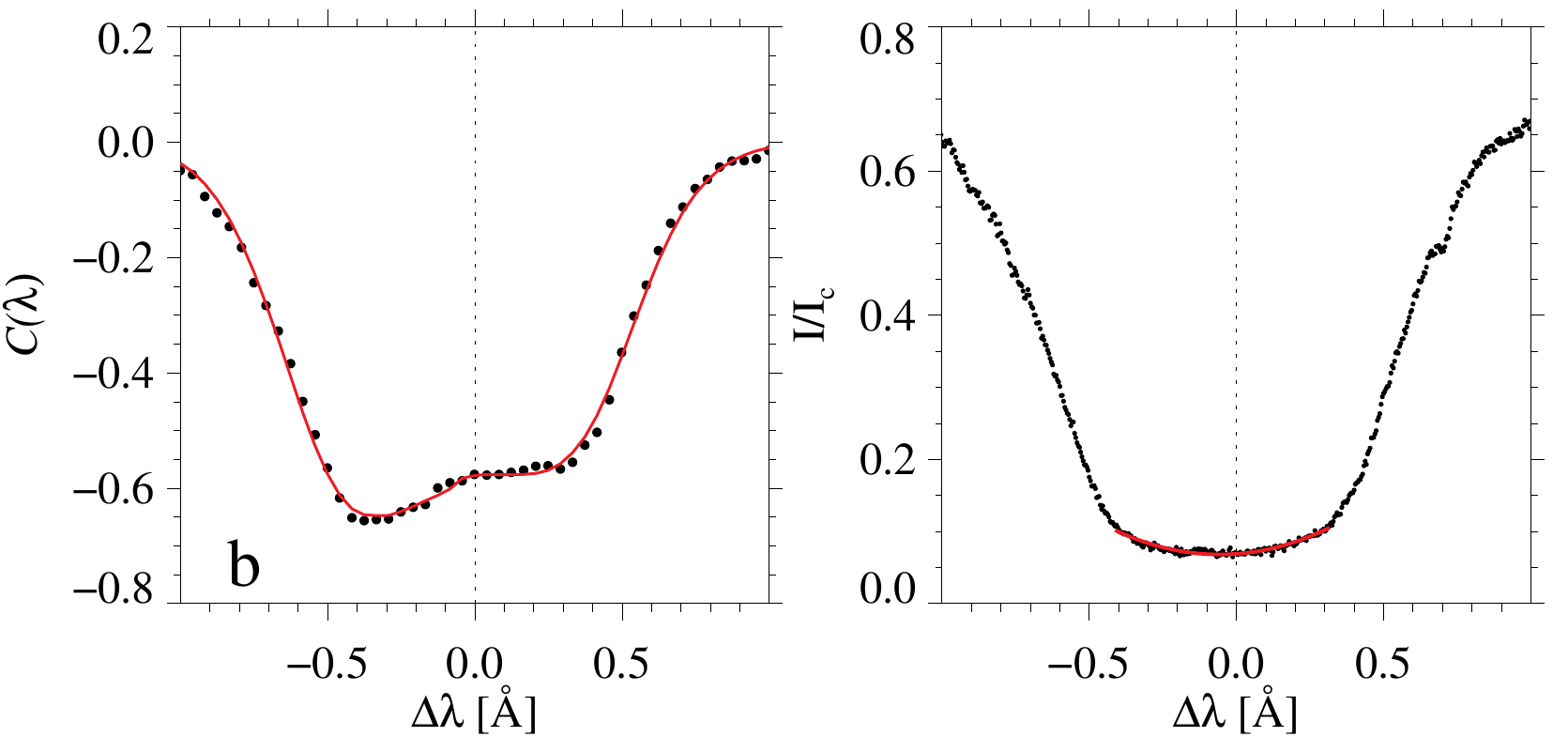}
\caption{The dots show the contrast profile (\textit{left}) and observed H$\alpha$ profile 
(\textit{right}). The profile belongs to the spine of the filament and is located at label `b' in Fig. 
\ref{Fig:vttmaps}. The red line shows the CM fit (\textit{left}) and the H$\alpha$ parabola line core fit 
(\textit{right}). The inferred velocities for the CM and line core fits are $-3.11$ and $-2.50$~km\,s$^{-1}$, 
respectively.  }
\label{Fig:contrasvsprof}
\end{figure}

As a comparison, we performed H$\alpha$ line core fits to infer the LOS velocities. The average velocities indicated 
that the filament, as a whole structure, had an upward motion towards higher layers of the atmosphere. This ascend was 
rather slow, on average $-0.225$~km\,s$^{-1}$, and was slightly more marked in the spine ($-0.288$~km\,s$^{-1}$), which 
is also perceptible in Fig.~\ref{Fig:vmaps}. 
However, this small generic upward trend could not be confirmed from the CM-inferred mean velocity, where the 
velocities were larger but on average canceled out yielding a value close to zero. Still, it is 
found that the velocity trend inferred from both methods, i.e., CM inversions versus 
line core fits, correlates well. 
As reported by \citet[][]{chae06}, 
we also confirmed that velocities inferred from CM inversions had larger values than the ones retrieved from line core 
fits (see Figs. \ref{Fig:velhistograms} and \ref{Fig:cloudhistograms}). This was also reflected in the standard 
deviation, which was a factor of $\sim$2 larger for the CM inferred velocities. 
Figure~\ref{Fig:contrasvsprof} shows an example of the different velocities inferred for the same 
pixel using the two different techniques. While the CM fit on the lefthand side of Fig.~\ref{Fig:contrasvsprof} 
yielded a velocity of $-3.11$~km\,s$^{-1}$, the H$\alpha$ line core fit on the righthand side provided a smaller 
velocity of $-2.50$~km\,s$^{-1}$. \citet[][]{chae06} ascribe this difference due 
to the fact that the line core fits provide information of the averaged line shifts between the underlying photosphere 
and the filament. The slow rise of the spine of the filament could be a precursor of a following 
eruption. A continuous rise of plasma might lead to a destabilization of the filament. Five days later, on 2011 
November 20 at around 16:00 UT, half of the filament lifted off as part of a CME observed with SDO/AIA. The fact that 
this huge filament underwent several eruptive events, before and after our data sets were recorded on 2011 
November 15, reveals the unstable nature of extremely large filaments. It is still surprising that the filament 
survived during the whole disk passage, disappearing on the western solar limb by the end of November 15.

A thorough inspection of the 
CM-inferred velocities along the filament (Fig.~\ref{Fig:cloudresults}) revealed that the velocity pattern is 
inhomogeneous, with different concentrations of up and downflows. This pattern was found throughout the filament, i.e., 
during the $\sim$100 min, which were necessary to scan the whole filament with the slit scanner. 
A close examination of barb `d' in both H$\alpha \pm 0.5$~\AA\ images revealed dark features which were only seen in 
the blue or red H$\alpha$ line wing image. The spatial separation between them was less than 1\arcsec. 
Thus, this indicated oppositely directed, i.e., counter-streaming, flows which could also be seen in the associated 
Doppler and CM velocity maps. In addition, \citet{diercke2014} found these counter-streaming flows in the spine of this 
filament by analyzing SDO/AIA time series. We are aware that, if this phenomenon occurs within the resolution element, 
the CM inversions will only provide an average of the physical parameters in that area. Hence, the lower than 
expected velocities for counter-streaming flows in barb `d' could be ascribed to not fully resolved 
threads with oppositely directed flows.

The \ion{Na}{i} D$_2$ inferred Doppler velocities were unaffected by the presence of the filament.  
Large conglomerates of brightenings matched magnetic field concentrations seen in the HMI magnetograms. Since the 
H$\alpha$ and \ion{Na}{i} D$_2$ data were simultaneously recorded at the VTT we could easily align the H$\alpha$ 
slit-reconstructed images with HMI. We detected mixed polarities at the end of barbs extending from the giant filament, 
as already reported for other smaller filaments in previous works 
\citep[e.g.,][]{martin98,chae05,lopezariste06,joshi13}. Nevertheless, this is not the case for all barbs of this giant 
filament (see Fig.~\ref{Fig:HaHMIoverlap}).

The global properties of horizontal proper motions for the quiet Sun and the 
photosphere below the filament were virtually the same. Only minute inspection of 
flow speed and divergence revealed flow kernels with a flow speed of 
0.30--0.45~km~s$^{-1}$ and a size of 5--8~Mm, in which magnetic flux was 
transported towards the filament, in particular at the ends of the filament. This finding might also indicate 
that the big gap seen on the lefthand side of the filament is being refilled by plasma, i.e., the 
filament is reestablishing its initial configuration.

Our study has revealed that extremely large quiescent filaments have similar properties as their smaller 
counterparts. Therefore, although this filament was a single structure when it appeared at the front 
side of the Sun, it could have been made out of junctions of smaller filaments as proposed by \citet[][]{anderson05}.
The origin and stability of such huge structures in the solar chromosphere remains a 
mystery and a 
challenge for the next generation of solar telescopes. The spatial resolution provided by SDO/HMI is insufficient to 
reveal any small details about the flux transport towards the filament.  Spectropolarimetric 
observations with high temporal and spatial resolution both in the photosphere 
and chromosphere are needed to elucidate how a filament forms along the filament 
channel. New multi-wavelength observations, e.g., with the GREGOR Fabry-P\'erot 
Interferometer \citep[GFPI,][]{denker10, puschmann12} and the Grating 
Infrared Spectrograph \citep[GRIS,][]{collados12} at GREGOR solar telescope 
\citep{schmidt12}, are a necessity to accomplish such challenging task.


\begin{acknowledgements}
The authors thank S. F. Martin, A. Warmuth, and A. Diercke for carefully reading the manuscript and for 
helpful comments. The authors also thank the anonymous referee for his/her thorough review and highly appreciate the
constructive comments on the paper. Discussions with P. Schwartz within the framework of the 
Deutscher Akademischer Austauschdienst (DAAD) project PPP Slowakei 
5706565721 'Das magnetische Vektorfeld solarer Filamente' are greatly acknowledged. 
C.K. thanks ISSI for enabling interesting discussions.
The Vacuum Tower Telescope and the Chromospheric Telescope are operated by the
Kiepenheuer Institute for Solar Physics in Freiburg, Germany, at the Spanish
Observatorio del Teide, Tenerife, Canary Islands. The ChroTel filtergraph has
been developed by the Kiepenheuer Institute in cooperation with the High
Altitude Observatory in Boulder, Colorado, U.S.A. NSO/Kitt Peak FTS data were
produced by NSF/NOAO.
H-alpha data were provided by the Kanzelh\"ohe Observatory, University of Graz, Austria.
M.V. acknowldeges support by SOLARNET, a project of the European Commission's FP7 Capacities 
Programme for the period April 2013 - March 2017 under Grant Agreement No. 312495. 
C.D. was supported by grant DE 787/3-1 of the German Science Foundation (DFG). 

\end{acknowledgements}


\bibliographystyle{aa}
\bibliography{biblio}

\end{document}